\newcommand{\al}{\alpha}  \newcommand{\ga}{\gamma}
\newcommand{\Om}{\Omega} \newcommand{\om}{\omega}

\newcommand{\ep}{\epsilon} \newcommand{\de}{\delta}
\newcommand{\R}{{\Bbb R}}\newcommand{\NNn}{{\Bbb N}}
\newcommand{\pa}{\partial}
 \newcommand{\G}{{\cal G}}
\newcommand{\DD}{{\cal D}}\newcommand{\X}{{\cal X}}
\newcommand{\bt}{\bar t}\newcommand{\bx}{\bar x}
 \newcommand{\br}{\bar r}
\newcommand{\nn}{\nonumber}\newcommand{\ba}{\begin{array}}
\newcommand{\ea}{\end{array}}\newcommand{\bea}{\begin{eqnarray}}
\newcommand{\bean}{\begin{eqnarray}}\newcommand{\eea}{\end{eqnarray}}
\newcommand{\beq}{ \begin{equation} }\newcommand{\beqn}{ \begin{equation*} }
\newcommand{\eeq}{ \end{equation} }\newcommand{\lb}{\label}
\newtheorem{proposition}{Proposition}
\documentstyle[a4,10pt,aps,prb]{revtex}
\input amssym.def\input amssym.tex
\title{\begin{flushleft} 
       {\small{\bf J. Math. Phys., Vol 38, No. 3, March 1997}}
        \end{flushleft}
\qquad\\
\qquad\\    
The ultrarelativistic Reissner-Nordstr\o m field in the Colombeau algebra}
\author{R. Steinbauer}
\address{Institute for Theoretical Physics, University of Vienna\\
       Boltzmanng. 5, A-1090 Wien/Vienna, Austria \\
       E-mail: stein@doppler.thp.univie.ac.at}
\begin{document}


\maketitle
\thispagestyle{empty}



\begin{abstract}
\quad\newline
\noindent
(Received 9 May 1996; accepted for publication 12 November 1996)
\vskip10pt
\noindent
The electromagnetic field of the ultrarelativistic Reissner-Nordstr\o m
solution shows the physically highly unsatisfactory property of a
vanishing field tensor but a nonzero, i.e., $\delta$-like, energy density.
The aim of this work is to analyze this situation from a mathematical
point of view, using the framework of Colombeau's theory of nonlinear
generalized functions. It is shown that the physically unsatisfactory 
situation
is mathematically perfectly defined and that one cannot avoid such
situations when dealing with distributional valued field tensors.
\end{abstract}

\vskip24pt


\section{INTRODUCTION}

Recently, there has been some interest in gravitational shock wave
geometries produced by ultrarelativistic particles.~\cite{bn1,bn2,dt}
In field and string theory one is interested especially in high energy
scattering processes in these geometries.~\cite{ls1,ls2,ls3}

In 1971 Aichelburg and Sexl~\cite{as} derived the
ultrarelativistic limit of the Schwarz\-schild metric, which turns out to be a
pp-wave with a distributional, i.\,e., $\de$-like profile function. This
result has been reproduced by several authors using various
methods,~\cite{dt,ls3} but all of them invoking
distributional techniques.  Loust\'o and S\'anchez~\cite{ls2}
derived the ultrarelativistic limit of the Reissner-Nordstr\o m solution. To
gain a distributionally well defined limit of the metric they had to rescale
the charge by $e^2=p_e^2(1-v^2)^{1/2}$ which also forces the limit of the
electromagentic field to vanish (even in the space of distributions).
However the limit of the energy stress tensor of the eletromagnetic field is
nonvanishing, thus leaving us with the physically highly unsatisfactory
situation of a vanishing field producing a $\de$-like energy density.

The aim of this work is to discuss this situation from a more mathematical
point of view. Computing the electromagnetic stress tensor from the field
tensor is a {\em nonlinear }operation and thus not defined within the
framework of classical distribution theory.  Thus, strictly speaking, if one
is dealing with distributional-valued field tensors one has to go beyond
distribution theory. In the 1980s\, J.F. Colombeau~\cite{c1,c2,c3}
developed a theory of generalized functions providing the possibility of a
product of distributions.  He constructed
differential algebras $\G$ containing the space of smooth functions as a
subalgebra and the space of distributions as a subspace. Thus Colombeau's
theory provides us with a natural framework for discussing situations like
the above-mentioned one from a more abstract viewpoint.~\cite{o,r} Moreover
most recently  first applications of this very framework to 
problems of classical field theories~\cite{gm} and general relativity~\cite{cl} 
have appeared.  

This work is organized in the following way. In Sec.~II we briefly
recall the basics of Colombeau's nonlinear theory of generalized functions.
To get
some working knowledge, in Sec.~III we take a look at the ultrarelativistic
limit of the Coulomb field in flat space, and finally in Sec.~IV we
discuss in some detail the electromagnetic field of the ultrarelativistic
Reissner-Nordstr\o m solution and its energy stress tensor.

 \section{BASICS OF COLOMBEAU THEORY}

Let $\Om$ be an open subset of $\R ^n$ or a smooth manifold.
We denote by $\DD(\Om)$ the space of
test functions on $\Om$, i.\,e., the space of smooth functions with compact
support in $\Om$, and by $\DD'(\Om)$ the space of distributions on $\Om$.
Finally the action of a distribution $w$ on a test fuction $\varphi$ we
denote by $\langle w,\varphi\rangle$.

In $\DD'$ no meaningful product can be defined. Moreover, as 
{\em L. Schwartz} showed in 1954, there even exists no associative and 
commutative
differential algebra containing the space of {\em continuous }functions as a
subalgebra that allows a linear embedding of the space of distributions in it.
{\em J.F. Colombeau}~\cite{c1,c2,c3} introduced differential algebras $\G$
containing the space of distributions as a subspace, and the space of {\em
smooth }functions as a faithful subalgebra, thus providing a natural
framework for studying nonlinear operations on singular data, i.\,e.,
distributions.

In this paper we consider the ``special variant'' of Colombeau's algebra, whose
definition we briefly recall. Set
\bea
{\cal E_M}(\Omega)
 &=& \{\,(u_\epsilon)_{\epsilon}\in(C^{\infty}(\Omega))^{(0,1)}:
 \,\forall K\subset\subset\Omega \, \forall \alpha\in \NNn_0^n\,\exists N>0:\nn
\\&&
\hskip3cm\sup\limits_{x\in K}\mid\partial^\alpha u_\epsilon(x)\mid=O
 (\epsilon^{-N})\quad(\epsilon\to 0)\,\}\,\,,\nopagebreak[0]
\\{\cal N}(\Omega)
 &=&\{\,(u_\epsilon)_{\epsilon}\in (C^\infty(\Omega))^{(0,1)}:
 \,\,\forall K\subset\subset\Omega \,\forall \alpha\in \NNn_0^n,\forall M>0:\nn
 \nopagebreak[0]
\\&&\hskip3cm \sup\limits_{x\in K}\mid\partial^\alpha u_\epsilon(x)\mid
  =O(\epsilon^{M})\quad(\epsilon\to 0)\,\}\,.
\eea
${\cal E_M}(\Om)$ is a differential algebra with pointwise operations and
${\cal N}(\Om)$ is an ideal in it. We define the {\em algebra of generalized
functions, }or {\em Colombeau algebra, }by \beq \G(\Om)\,:=\,{\cal
E_M}(\Om)\,/\,{\cal N}(\Om)\eeq and denote its elements by
$$ u\,=\,(u_\ep)_\ep\,+\,{\cal N}(\Om)\,.$$

Distributions with compact support on $\R^n$ can now be embedded into
$\G(\R^n)$ by convolution with a {\em mollifier $\rho_\ep$, }defined as
follows; let $\rho\in {\cal S}(\R^n)$ (Schwartz's space) with the properties
$\int\rho(x)\,dx=1$ and $\int x^\alpha\rho(x)\,dx=0 \quad\forall \alpha\in
\NNn^n,\,\mid\!\al\!\mid\geq1$, then we set
$\rho_\ep(x):=(1/\ep^n)\rho(x/\ep)$. So we have the map
$\iota_0(\om)=(\om*\rho_\ep)_\ep+{\cal N}(\R^n)$.\newline
This embedding can be
``lifted'' to an embedding $\iota: $ $\DD'(\Om)\hookrightarrow\G(\Om)$ by means
of sheaf theory while smooth functions are embedded as constant sequences,
i.\,e. $\sigma(f)=(f)_\ep$.

Next we briefly recall the concept of association in the Colombeau algebra,
of which we make extensive use in the physical calculations of
Secs.~III and IV.  One particular property of Colombeau's
algebra is the so-called ``coupled-calculus,'' namely equality in the algebra
of generalized functions and equality on distribution level.  The latter is
defined as follows: a generalized function $u$ is called {\em associated to
$0$} ($u\approx 0$) if for one (and hence any) representative $(u_\ep)_\ep$
we have
\beq
 \lim\limits_{\ep\to 0}\int u_\ep(x)\varphi(x)\,dx\,=
 \,0\,,\qquad\forall\varphi\in\DD(\Om).
\eeq

Association defines an equivalence relation in the Colombeau algebra that
clearly is coarser than equality and is compatible with differentiation,
i.\,e., $u\approx v\Rightarrow \pa^\al u\approx\pa^\al v$ but is not
compatible with the algebra multiplication, i.\,e., $u\approx
v\not\Rightarrow wu\approx wv$.

If a generalized function $u$ is associated to $\iota(w)$, where $w$ is a
distribution, then one says that $w$ is the {\em distributional shadow }or
the {\em associated distribution }of $u$. Not all Colombeau functions have a
shadow-for example $\de^2$, i.e., $\iota(\de)^2$-but if
a generalized function has a shadow
the latter is unique. Different generalized functions may have the same
shadow, providing us with the possibility to model singular situations in
various ways.~\cite{c3}

\section{THE ULTRARELATIVISTIC LIMIT OF THE COULOMB FIELD}\lb{p1}

In Cartesian coordinates the Coulomb field is given by the Maxwell field
tensor components $F^{0\al}=-(ex^\al)/r^3$ where $r^2=x^2+y^2+z^2$ and
$\al=1,2,3$, and the other components vanishing.
We apply a Lorentz boost with velocity $v$ in the $x$ direction
and, in order to simplify computations, we additionally transform to a frame
associated with null coordinates; in particular
\bea
 t\,\mapsto\,\bt=\ga(t+vx)&\mapsto &u=\bx-\bt\,,\nn
\\x\,\mapsto\,\bx=\ga(x+vt)&\mapsto &w=\bx+\bt\,,
\eea
where $\ga:=(1-v^2)^{-1/2}$.  Thus the boosted field tensor takes the form
{\normalsize $$
 F^{ik}=\frac{e(1-v^2)}{[(\bx-v\bt)^2+(1-v^2)\rho^2]^{3/2}}\left(\ba{cccc}
 0&2(\bx-v\bt)&(1-v)y&(1-v)z\\ -2(\bx-v\bt)&0&-(1+v)y&-(1+v)z
\\-(1-v)y&(1+v)y&0&0\\ -(1-v)z&(1+v)z&0&0 \ea\right),
$$}\noindent
where $\rho^2=y^2+z^2$.

We could now view the components of the field
tensor as sequences of locally integrable functions parametrized by the
boost velocity $v$, and calculate their limits in the space of distributions
as $v\to 1$. However, it is well known that in this limit some
components of the field tensor tend to $\de(u)$ and hence the energy
momentum tensor $T_{ik}=F_{il}F^l\!_k-1/4 F_{ml}F^{lm}$ of the
ultrarelativistic field cannot be calculated directly due to the lack of a
multiplication in the space of distributions. Alternatively one could think
of first computing the energy momentum tensor of the boosted field and then
taking the limit. However, in this approach some components of the momentum
tensor diverge in the space of distributions, thus making it again
impossible to compute the energy momentum tensor of the ultrarelativistic
Coulomb field within the framework of classical distribution theory.  

To overcome these difficulties we embed the components $(F^{ik}\,_v)_v$ of
the field tensor into the Colombeau algebra. The physically most natural way
to do so is to view a whole sequence (for example $(F^{01}\,_v)_v$) as {\em
one }Colombeau function. Generalized functions in the sense of Colombeau are
classes of sequences of {\em smooth }functions, but our sequences are not
even defined on the set ${\cal A}=\R^4\setminus\{x=vt\,(0<v<1)\,
\wedge\rho=0\}$, where from now on we neglect the bars over $x$ and $t$ to
simplify the notation. So in order to get an ``embedding'' of $(F^{ik}\,_v)_v$
into Colombeau's algebra without cutting the whole set ${\cal A}$ from the
domain we employ the following construction that only forces us to restrict
the domain to $\Om:=\R^4\setminus \{x=t\wedge\rho=0\}$, which is physically
reasonable since we precisely cut out the world line of the
ultrarelativistic particle.  Let $\X_v\in C^\infty(\R)\,\,(0<v<1)$ be such
that
$$\begin{array}{rcl}
 \X_v(\xi)\equiv 1\,,&&\mid\xi\mid\geq 4(1-v)^2\,,\\
 \X_v(\xi)\equiv 0\,,&&\mid\xi\mid\leq \,(1-v)^2\,\,.
\end{array}$$
Using the abbreviation $a:=(x+vt)^2+\rho^2$ we define
\beq
 \tilde F^{ik}\,_v\,=\,\left\{\begin{array}{ll}F^{ik}\,_v&a>4(1-v)^2
\\F^{ik}\,_v\,\X_v(a)\qquad&4(1-v)^2\geq a\geq (1-v)^2
\\0&(1-v)^2>a\,,\end{array}\right.
\eeq
which lies in $\G(\Om)$ and equals $F^{ik}\,_v$ outside a cylinder of radius
$2(1-v)$ with axis $x=vt$ in the $\rho=0$ plane, whereas it is smoothed down
to zero inside a cylinder with half this radius. Note however that $\tilde
F^{ik}\,_v$ is smooth even on $\R^4$. Now $\tilde F^{ik}\,_v$ is the best
possible approximation for $F^{ik}\,_v$ in $\G(\Om)$ in the following
precise sense: Given any compact $K\subset\Om$, finally (i.e., $v$ large
enough) $\tilde F^{ik}\,_v$ equals $F^{ik}\,_v$ on an open neighborhood of
$K$.

Putting $v(\ep)=1-\ep$ we get the desired ``embedding'' of the field tensor
into the Colombeau algebra
$$
 (F^{ik}\,_v)_v\,\mapsto\,(\tilde F^{ik}_{\,v(\ep)})_\ep+{\cal N}(\Om).
$$

Now we are able both to compute the energy momentum tensor in $\G(\Om)$, and
to make use of the concept of association, which corresponds physically to
taking of the ultrarelativistic limit. First we calculate the association
relations for the field tensor.
\begin{proposition}\quad
 \beq
 \tilde F^{ik}\,\approx\,
 \left(\begin{array}{cccc} 0&0&0&0
  \\ 0&0&-y&-z \\ 0&y &0&0 \\ 0&z &0&0
 \end{array}\right)
 \frac{4e}{\rho^2}\,\delta (u)\quad
 \eeq
\end{proposition}
\vskip6pt

\noindent{\em Proof: }
We have to compute the following limits \beq\lim\limits_{\ep\to
0}\int\limits_\Om \tilde F^{ik}_{\,v(\ep)}(t,x,y,z)\, \varphi(t,x,y,z)\,
dtdxdydz\,\,, \eeq where $\varphi$ is an arbitrary test function in
$\DD(\Om)$. We only carry out the calculations for $\tilde F^{01}$ and
$\tilde F^{21}$ as the other components are either related to these by
symmetry or the limits can be taken in a closely analogous way. We choose
$v$ so large that on an open neighborhood of the support of $\varphi$ we
have $\tilde F^{ik}_v=F^{ik}\,_v$, and for simplicity we set
$[\quad]:=(x-vt)^2+(1-v^2)\rho^2$. Then we have
\bea
 \left|\int_{\mbox{supp}(\varphi)}\tilde F^{01}_{\,v(\ep)}
  \,\varphi\right|
 &\leq&2e\|\varphi\|_\infty\int\int\int\left[\int_{-\infty}^\infty
 \left|\frac{(1-v^2)\,(x-vt)}{[\quad]^{3/2}}\right| dx\right]dtdydz \nn
\\&=&8\pi e\sqrt{1-v^2}\|\varphi\|_\infty \int\int
  \frac{1}{\rho}\,\rho d\rho dt\,\,,
\eea
which vanishes in the limit, since the remaining integral has to be taken
over a compact set only.\newline
To prove $\tilde F^{21}\,_{v(\ep)}\approx\,
(4ey/\rho^2)\de(u)$ we first note that for $\tilde f_{v(\ep)}$,
defined from $f_v=ey(1+v)(x-vt)/(\rho^2[\quad]^{1/2})$ in the same way as
$\tilde F^{ik}\,_{v(\ep)}$ is defined from $F^{ik}\,_v$ we have
\beq
 \lim\limits_{\ep\to 0}\int\limits_{\mbox{supp}(\varphi)}\tilde 
 f_{v(\ep)}\varphi
 \,=\,\left\langle\,\frac{2ey}{\rho^2}[\theta(x-t)-\theta(t-x)],\varphi\,
     \right\rangle\,\,,
\eeq
where we have used Lebesgue's dominated convergence. Differentiating this
relation with respect to $x$ we get the desired result. $\qquad \Box$
\vskip12pt

Next we compute the energy stress tensor in $\G(\Om)$ by componentwise
multiplication to get
{\scriptsize
\bea
  \tilde T_{ik\,v(\ep)}&=&
  \frac{e^2(1-v^2)^2}{2((x-v t)^2+(1-v^2)\rho^2)^3}
 \nn\\\quad\nn\\&&\times  
 \left(\begin{array}{cccc}
    \frac{1}{2}(1+v)^2\rho^2  &\frac{1}{2}( x-v t)^2
    &(1+v)( x-v t)y
    &(1+v)( x-v t)z  \\ \\
    \frac{1}{2}( x-v t)^2  &\frac{1}{2}(1-v)^2\rho^2
    &-(1-v)( x-v t)y
    &-(1-v)( x-v t)z  \\ \\
    (1+v)( x-v t)y
    &-(1-v)( x-v t)y
    &( x-v t)^2+  &-2(1-v^2)yz \\
    &&+(1-v^2)(z^2-y^2) \\ \\
    (1+v)( x-v t)z &-(1-v)( x-v t)z
    &-2(1-v^2)yz  &( x -v t)^2+ \\
    &&&+(1-v^2)(y^2-z^2)
  \end{array}\right)
\eea
}\noindent for $a>4(1-v)^2$ and 
smoothed down to zero ``inside'' in the same way as in
the case of $\tilde F^{ik}\,_{v(\ep)}$.

\begin{proposition} \qquad
\bea
\tilde T_{ik}&\approx&0\quad\forall (i,k)\not=(0,0) \nn
\\ \tilde T_{00}&&\mbox{{\em has no associated distribution.}}
\eea\end{proposition}

\noindent{\em Proof: }
The calculation for all the components is closely analogous to the one given
above with the exception of $T_{00}$, so we shall only examine this one in
detail. As a test function $\varphi$ we take $1/\rho^2$ on the set $x,t\in
[-N,N],\,\phi\in [0,2\pi]$ and $\rho\in [r_0,r_1]$ $(0<r_0)$, where $\phi$
is the polar angle in the $(y,z)$-plane, and let it smoothly approach zero
``outside''. Thus we have for large enough $v$ 
\bea
 \int\limits_{\mbox{supp}(\varphi)}\tilde T_{00\,v(\ep)}\,\varphi
 &\geq&\frac{e^2(1+v)^2\,(1-v^2)^2}{4}\int\limits_{-N}^N \int\limits_{-N}^N
 \int\limits_0^{2\pi}\int\limits_{r_0}^{r_1}\frac{\rho^2}{[\quad]^3}
 \frac{1}{\rho^2}\,\rho d\rho d\phi dxdt \nn
\\&=&\frac{e^2\pi (1+v)^2(x-vt)}{16\,v\,\rho^3\sqrt{1-v^2}}
 \,\arctan\frac{x-vt}{\sqrt{1-v^2}\rho}\mid_{r_0,-N,-N}^{r_1,N,N} \nn
\\&&+\frac{e^2\pi(1+v)^2}{32\,v\,\rho^2}\ln(\,[\quad]\,)\mid_{r_0,-N,-N}^
 {r_1,N,N}\nn
\\&&-\frac{e^2\pi(1+v)^2}{32\,v\,\rho^2}\ln\left(1+\frac{(x-v t)^2}{(1-v^2)
 \rho^2}\right)\mid_{r_0,-N,-N}^{r_1,N,N}\,\,.
\eea
In the limit the second term diverges whereas the other two terms vanish.
$\quad\Box$
\vskip12pt

We see that computing the energy momentum tensor we fall out of the class
of Colombeau functions with associated distributions.
Taking the square of the two functions $(F^{12}_{v(\ep)})_\ep$ and
$(F^{13}_{v(\ep)})_\ep$, both of them associated to the $\de$ distribution,
we get the generalized function $(T_{00\,v(\ep)})_\ep=
-(1/4)\,\Bigl((F^{12}_{v(\epsilon)})^2+(F^{13}_{v(\epsilon)})^2
\Bigr)_\epsilon$ not allowing any associated distribution, showing that
$(T_{00\,v(\ep)})_\ep$ is an object {\em only } defined in the Colombeau
algebra.  Due to the properties of Colombeau's theory one could subject it
to a wide class of nonlinear operations and even use it as a source term for
nonlinear PDEs and then check if the resulting generalized function has an
associated distribution.

Of course this result mirrors
the fact that if we had computed the energy tensor ``naively'' from the
ultrarelativistic limit of the field tensor in $\DD'$ we would have obtained
$T_{00}=(4e^2/\rho)$``$\de^2(u)$,'' all other components vanishing. This
suggests a ``commutativity'' of the limiting procedure and ``multiplication,''
i.\,e., the computation of the stress tensor from the field tensor. However,
we again point out that no meaningful product can be defined within the
framework of distribution theory but only in the Colombeau algebra. As a 
limiting procedure we have used the concept of association in $\G(\Om)$,
which is not compatible with the product, and so the above-mentioned
``commutativity'' does not hold in general, as we are going to see in the next
section.

\section{THE ULTRARELATIVISTIC LIMIT OF THE REISSNER-NORDSTR\O M
         SOLUTION}\lb{p2}

The RN metric representing the gravitational field of a point particle with
mass $m$ and charge $e$ in isotropic coordinates is given by
\beq  ds^2\,=\,(1-\frac{2m}{r(\br)}+\frac{e^2}{r^2(\br)})\,dt^2
-(1+\frac{m}{\br}+\frac{m^2-e^2}{4\br^2})^2\,(d\br^2+d\Om^2)\,,
\eeq
where $r=\br[1+m/\br+(m^2-e^2)/(4\br^2)]$ is the radial
coordinate in a Schwarzschild-like coordinate system. We apply a boost in
$x$ direction relative to an asymptotic, Cartesian coordinate system
associated with the isotropic radius $\br$. To obtain a distributionally well
defined result in the ultrarelativistic limit we have to rescale
the mass and the charge~\cite{ls2} in the following manner:
$$ m\,=\,p/\ga,\qquad e^2\,=\,p_e^2/\ga,$$
where in the limit we keep the momenta $p$ and $p_e$ fixed. The rescaling of
the mass is well motivated~\cite{as} and saves the energy of the
particle from diverging due to its finite rest mass by keeping the total
energy $p$ fixed and letting $m$ approach zero in the ultrarelativistic
limit. The rescaling of the charge is not too well motivated. However, the
limit of the metric is given by~\cite{ls2}
\beq ds^2\,=\,\Bigl\{8p\ln\rho+\frac{3}{2}\pi\frac{p_e^2}{\rho}\Bigr\}
\,\delta(u)\,du^2-dudw-dy^2-dz^2\,\,,
\eeq
\noindent
where $u$ and $w$ again denote null coordinates defined as in the last
Section.
This metric belongs to the class of pp waves and is flat everywhere except
on the null plane $u=0$ which is normal to the boost direction and contains
the particle.

However, the rescaling of the charge dramatically acts on the limit of the
electromagentical field and its stress energy tensor. In a frame associated
with null coordinates the boosted field tensor takes the form
$$ F_{ik}\,=\,\frac{e\,(1-v^2)
   G(\bar r)}
   {((x-v t)^2+(1-v^2)\rho^2)^{3/2}}
   \left(\begin{array}{cccc}0& x-v t& y& z \\
                            - x+v t&0&-v y&-v z\\
                            - y&v y&0&0 \\
                            - z&v z&0&0
   \end{array}\right)\,, $$
where the factor $G(\bar r)=4\bar
r^2(4\bar r^2-m^2+e^2)/(4\bar r^2+4\bar r m+m^2-e^2)^2$ arises from the
transformation to isotropic coordinates and we again have neglected the bars
over $x$ and $t$. Now we embed $(F_{ik\,v})_v$ into the Colombeau algebra in
the same manner as in the Coulomb case, again denoting the
$\G(\Om)$ function best approximating $F_{ik}$ by $(\tilde
F_{ik\,v(\ep)})_\ep$. By calculations very similar to those carried out
in Sec.~II and taking into account the rescaling of mass and charge we get
the following association relations.

\begin{proposition}\qquad
\beq \tilde F_{ik}\,\approx\,0\,\qquad\forall i,k\,.\eeq
\end{proposition}

Next we compute the stress energy tensor. For $a>4(1-v)^2$ we have
\beq\tilde T_{ik}\,=\,\frac{p_e^2G^2(\bar r)}{((x-vt)^2+(1-v^2)\rho^2)^3}\eeq
\vskip12pt
{\scriptsize
 \begin{flushleft}
  $\left(
  \begin{array}{cccc}
       (1-v)^{3/2}(1+v)^{7/2}(g^{11}-g^{00})( x-v t)^2/4
      &(1-v^2)^{5/2}(g^{00}+g^{11})( x-vt)^2/4
    \\
       +(1-v)^{5/2}(1+v)^{9/2}(g^{00}+g^{11})\rho^2/4
      &+(1-v^2)^{7/2}(g^{11}-g^{00})\rho^2/4
    \\ \\
       (1-v^2)^{5/2}(g^{00}+g^{11})( x-vt)^2/4
      &(1-v)^{7/2}(1+v)^{3/2}(g^{11}-g^{00})( x-v t)^2/4
    \\
       +(1-v^2)^{7/2}(g^{11}-g^{00})\rho^2/4      
      &+(1-v)^{9/2}(1+v)^{5/2}(g^{00}+g^{11})\rho^2/4
    \\ \\
       (1-v)^{5/2}(1+v)^{7/2}g^{00}( x-v t) y
      &-(1-v)^{7/2}(1+v)^{5/2}g^{00}( x-v t) y   
  \\ \\
      (1-v)^{5/2}(1+v)^{7/2}g^{00}( x-v t)z
      &-(1-v)^{7/2}(1+v)^{5/2}g^{00}( x-v t) z
  \end{array}\right.
  $
  \end{flushleft}
  \vskip18pt
  \begin{flushright}
  $
  \left.\begin{array}{cc}
      (1-v)^{5/2}(1+v)^{7/2}g^{00}( x-v t) y
      &(1-v)^{5/2}(1+v)^{7/2}g^{00}(x-vt) z
    \\ \\
      -(1-v)^{7/2}(1+v)^{5/2}g^{00}( x-vt)y
      &-(1-v)^{7/2}(1+v)^{5/2}g^{00}( x-v t) z
    \\ \\
      (1-v^2)^{5/2}g^{00}
               [( x-v t)^2+(1-v^2)( z^2- y^2)]
      &-2(1-v^2)^{7/2}g^{00} y z
    \\ \\
      -2(1-v^2)^{7/2}g^{00} y z
      &(1-v^2)^{5/2}g^{00}
               [( x-v t)^2+(1-v^2)( y^2- z^2)]
    \end{array}\right),
  $
 \end{flushright}
}
\vskip12pt
\noindent
where 
\beq g^{00}(\bar r)=(1+\frac{m}{\bar r}+\frac{m^2-e^2}{4\bar r^2})^{2}
(1+\frac{e^2-m^2}{4\bar r^2})^{-2}
\eeq 
and 
\beq g^{11}(\bar r)=(1+\frac{m}
{\bar r}+\frac{m^2-e^2}{4 \bar r^2})^{-2}
\eeq
and $\tilde T_{ik}$ smoothed down to zero ``inside'' again in the same way.
By a rather lengthy calculation, expanding $G(\bar r)$, $g^{00}$ and
$g^{11}$ and again taking into account the rescaling we get the following.

\begin{proposition}\qquad
\bea     T_{ik}&\approx&0\,\,,\qquad\forall\,(i,k)\not=(0,0)\,, \nn\\
         T_{00}&\approx&\frac{3p_e^2}{16\rho^3}\,\delta(u)\,.
\eea
\end{proposition}

\section{CONCLUSION}

The fact that the ``square'' $T_{00}$ of the generalized functions 
$(F_{ik})_\ep$
associated to $0$ is not associated to $0$ but to the $\de$ distribution is
not surprising from the viewpoint of Colombeau's theory. Association is
not compatible with multiplication in the algebra of generalized
functions. Thus we can say that the physically unsatisfactory situation of a
vanishing field but nonzero stress tensor is mathematically perfectly
defined. We see the incompatibility of linear idealisations like the $\de$
distribution and nonlinear computations. Colombeau's theory clearly cannot
solve this principal defect but provides us with a language and framework that
makes it possible to discuss such situations.

The question of the physical relevance of a zero field producing a
nonvanishing energy density clearly has to remain open. But from the
mathematical viewpoint such a situation cannot be avoided if one is dealing
with distributionally shaped fields.

\section*{ACKNOWLEDGMENTS}
Work supported by Doktorandenstipendium der
\"Osterreichischen Akademie der Wissenschaften, \# 338.

This work is part of my thesis~\cite{s} supervised
by H. Urbantke, whom I have to thank for his extensive help.
Also I want to thank M. Grosser, G. H\"ormann and M. Kunzinger for
many helpful discussions and M. Oberguggenberger for the critical reading
of the manuscript. 


\end{document}